\documentclass[10pt,journal,compsoc]{IEEEtran}

\usepackage{graphics}
\usepackage{subcaption} 
\usepackage{xcolor}
\usepackage{ulem} 
\usepackage{dirtytalk}
\usepackage{graphicx}
\usepackage{hyperref}
\ifCLASSOPTIONcompsoc
  \usepackage[nocompress]{cite}
\else
  \usepackage{cite}
\fi
\ifCLASSINFOpdf
\else
\fi
\begin{document}

\title{Contradicted by the Brain: Predicting Individual and Group Preferences via Brain-Computer Interfacing}

\author{Keith~M.~Davis~III,
        Michiel~Spapé,
        and~Tuukka~Ruotsalo% <-this % stops a space
\IEEEcompsocitemizethanks{\IEEEcompsocthanksitem Keith M. Davis III is with the Department
of Computer Science, University of Helsinki, Finland.
\protect\\
E-mail: keith.davis@helsinki.fi
\IEEEcompsocthanksitem Michiel Spapé is with the Department
of Psychology and Logopedics, University of Helsinki, Finland.% <-this % stops an unwanted space
\IEEEcompsocthanksitem Tuukka Ruotsalo is with the Department of Computer Science, University of Helsinki, Finland and the Department of Computer Science, University of Copenhagen, Denmark. \protect\\
E-mail: tr@di.ku.dk
}
\thanks{Manuscript received 10 May 2022; revised 14 October 2022; accepted 22 November 2022. Date of publication 15 December 2022; date of current version 29 November 2023.
This work was supported by the Academy of Finland and the Horizon 2020 FET program of the European Union through the ERA-NET Cofund funding under grant CHIST-ERA-20-BCI-001.
This work involved human subjects or animals in its research. Approval of all ethical and experimental procedures and protocols was granted by the University of Helsinki Ethical Review board and performed in line with the Declaration of Helsinki.
(Corresponding author: Keith M. Davis III.)
Digital Object Identifier no. 10.1109/TAFFC.2022.3225885}}


\markboth{Appears in IEEE TRANSACTIONS ON AFFECTIVE COMPUTING, VOL. 14, NO. 4, OCTOBER-DECEMBER~2023, DOI 10.1109/TAFFC.2022.322588510}%
{Davis \MakeLowercase{\textit{et al.}}: Contradicted by the Brain: Predicting Individual and Group Preferences via Brain-Computer Interfacing}



\IEEEtitleabstractindextext{%
\begin{abstract}
We investigate inferring individual preferences and the contradiction of individual preferences with group preferences through direct measurement of the brain. We report an experiment where brain activity collected from 31 participants produced in response to viewing images is associated with their self-reported preferences. First, we show that brain responses present a graded response to preferences, and that brain responses alone can be used to train classifiers that reliably estimate preferences. Second, we show that brain responses reveal additional preference information that correlates with group preference, even when participants self-reported having no such preference. Our analysis of brain responses carries significant implications for researchers in general, as it suggests an individual's explicit preferences are not always aligned with the preferences inferred from their brain responses. These findings call into question the reliability of explicit and behavioral signals. They also imply that additional, multimodal sources of information may be necessary to infer reliable preference information.
\end{abstract}

}

\maketitle

\IEEEdisplaynontitleabstractindextext

% \fi
\IEEEpeerreviewmaketitle

\section{Introduction}
\IEEEPARstart{R}{ecords} of user interactions with digital content are a critical component of user modeling and are utilized to better understand both users and content. This understanding informs a variety of digital applications, from tailoring advertisements \cite{alt2012increasing} and personalization of search results \cite{hannak2013measuring, bennett2012modeling} to guiding the creation of new content \cite{heck2021subconscious}. It has been widely demonstrated that behavioral user signals, such as clicks, ratings, or dwell time, can have a high utility for adaptation and personalization of information \cite{Teevan:2005:PSV:1076034.1076111,Joachims:2007}. Despite this success, the pitfall of behavioral signals is that they are only available for content that users directly interact with, such as viewing, rating, or sharing multimedia. Users may encounter content they personally prefer, but if they do not engage with it, their perceptions of the content are not captured through their expressed behavioral signals.

An intriguing alternative to user behavior is to directly obtain continuous measurements related to user cognition while they perceive information. Researchers have utilized physiological responses measured from human peripheral physiology \cite{Barral:2015:EPP:2678025.2701389,barral2016extracting} and brain signals to study neural correlates of perceived information \cite{Kangassalo:2019:WUI:3331184.3331243}, detect and predict relevance \cite{Gwizdka:2017,Allegretti:2015:RJH:2766462.2767811,Eugster:2014:PTB:2600428.2609594,Moshfeghi:2013UNDERS}, and even use as feedback in information access systems \cite{EugsterSREP,Moshfeghi:2013,Arapakis:2008:AFI:1390334.1390403}.

Previous work, however, has been limited to studying objectively defined relevance --- that is, using physiological and neural signals to detect whether the presented information is associated with a particular topic primed as relevant for the user prior to an experiment. Then, the evoked brain potentials are associated with information that either matches a predefined task (relevant) or information that violates that task (irrelevant). For example, in \cite{Eugster:2014:PTB:2600428.2609594} participants were asked to concentrate on a single topic, such as "cat". The relevance of associated individual terms (e.g. "feline" or "furry") was successfully predicted to be relevant. Similarly, in \cite{pinkosova2020cortical}, the graded relevance of presented text answers to prompted questions was associated with the corresponding neurophysiological responses, while in \cite{10.1145/1646396.1646433}, participants' emotional reactions were associated with the relevance of video content. However, the nature of how users experience content is far more nuanced.

For decades, it has been understood that users are not only evaluating topical associations but that their preferences are subjective and may depend on cognitive and affective processes that are more complex than mental categorization \cite{10.1002/asi.10137}. Consider a user browsing a video streaming service who is searching for a horror film to watch. In the context of relevance assessment, any horror film the streaming service recommends will be considered topically relevant, while films of other genres will not be considered relevant. However, whether or not a given film was \textit{enjoyed} by the user is distinct from relevance, as it depends on a personal preference assessment which is influenced by a variety of factors beyond genre and varies significantly depending on the individual. Estimating these preferences currently requires data on how users interact with content. Various methods have been proposed to infer preference information from behavioral signals, such as clicks or dwell time. Yet such signals are, at best, probes for the real preferences experienced by the users. 

Here, we study the phenomena of brain responses and their association with individual and group preferences. Individual preferences are defined as the preferences (what an individual likes or otherwise responds positively to) expressed by a single person, while group preferences are defined as aggregate preferences across the same peer group - that is, items that most members of the group like or respond positively to. Past work has shown that personal relevance can be predicted using brain responses. However, there is limited research on how reliably brain responses correlate with the group preference and explicitly expressed preferences. While an individual may explicitly report conscious preferences that are reflected in their brain responses, their brain responses may also contain information about social or cultural influences that are implicit in nature and otherwise outside of the individual's immediate awareness.

To study the association between brain responses, explicit self-reported preferences, and group preferences, we ask the following research questions:

\begin{enumerate}

\item[] {\bf RQ1:} Can brain responses be used to estimate graded user preferences?

\item[] {\bf RQ2:} Can group preferences be estimated from an individual's brain responses?

\end{enumerate}

First, we report a neurophysiological experiment where participants' natural preference reactions towards facial stimuli were recorded. Using this data, we show that explicit brain responses reflect graded preference judgments. Graded preferences refer to preferences that are assessed on a scale, in our case from 0 to 3, with 0 indicating "no preference" and 3 indicating "highest preference." Moreover, we show that brain responses can conflict with participants' preferences communicated via explicit self-report and that these brain responses correlate with group preference --- that is, preferences that are expressed by most members of a peer group.

Second, we report two machine learning analyses. In the first analysis, we train a model to classify graded user preferences from data collected via electroencephalography (EEG) and show that it is indeed possible to classify graded preferences from brain responses. In the second analysis, we show that additional preference information can be revealed by matching individual brain responses to the group preference. That is, even in cases when an individual did not explicitly report a preference, brain responses to stimuli that were highly ranked among their peers could be distinguished from brain responses to stimuli that had a low preference rank, and this effect was large enough to be reliably classified for half of the participants.

In summary, our contributions are the following:

\begin{enumerate}
    \item We show that explicit brain responses captured via EEG reveal information that corresponds to self-reported preferences, and that the decoding of different levels of user preferences is possible in a single-trial machine learning setting.
    
    \item  We show that brain responses can reveal additional information that is not otherwise captured from self-reported ratings, and which correlates with the group preference. For approximately half of the individuals, their brain responses followed the group preference and \textit{directly contradicted} their explicitly reported preferences.
    
\end{enumerate} 

\section{Background} 

Our work draws from the fields of human-computer interaction, psychology, and neurophysiology. We provide a short discussion of current methods in human-computer interaction to estimate user preferences from observable signals. Following, we explore the current state of the art with regards to user signals derived from human electrophysiology, such as through electroencephalography (EEG), -myography (EMG), -cardiography (EKG), and electrodermal activity (EDA). We also provide a brief overview of alternative methods like functional near-infrared spectroscopy (fNIRS), which offers unique benefits not found in EEG and is growing in popularity as a signal source for brain-computer interfaces (BCIs).

\subsection{Estimating preferences from human-computer interaction}

A wide variety of approaches and computational methods have been developed to model and predict user preferences from behavioral signals, such as dwell time \cite{Yi:2014,Kim:2014:MDT:2556195.2556220}, clicks and input reformulations \cite{Joachims:2007}, gaze and mouse tracking \cite{Guo:2012}, and touch interactions \cite{Guo:2013}. These signals can be roughly divided into two categories based on the information sources they rely upon: explicit interactions and implicit interactions. Explicit interactions refer to information that a user inputs to a preference management system on purpose. For example, a rating, binary expression of liking or disliking content, or categorical information on a user's content tastes \cite{burke2002hybrid}, can all be categorized as explicit interactions.

Implicit interactions, sometimes called implicit feedback, refer to different sorts of user behavior, such as clicks, watching habits, browsing activity, or other interactions that serve as indirect indicators of user preferences. Unlike explicit interactions, implicit interactions alone do not provide direct input from the users regarding their preferences.

Convincing empirical evidence of utilizing implicit feedback has been provided to support the performance of a variety of different implicit methods \cite{Ghorab2013,Kofler:2016}. However, while implicit behavioral signals have often been found effective, they are still at best only \textit{proxies} of the underlying preferences. Users may visit or spend time on certain content for reasons other than personal preference, for example simply out of curiosity. In cases of multimedia content such as images, movies, or videos, browsing and search, consumption behaviors like watch time or dwell time \cite{covington2016deep} are typically used to evaluate preferences and may lead to high or low preference estimations depending on the particular user's habits.

Notably from the perspective of the present work, we argue that all of these preference signals are based on a user's \textit{explicit} interactions. Even clicks or dwell time are based on a user explicitly interacting with some content, while the preference toward that content may be \textit{implicitly} expressed via this interaction. Here, we aim to capture implicit preferences that are not manifested from direct and explicit physical interactions but are observed via brain-computer interfacing. That is, the only interaction data used are evoked brain potentials produced from visual perception. Moreover, we aim to capture truly implicit preferences -- preference information that users may not be aware of, but that are still observable from their brain responses.

\subsection{Estimating preferences from physiology}

In recent years, physiological methods for evaluating user preferences have become objects of investigation with the increased availability of technology and improved methodology of analysis. In general, psychophysiological methods for detecting preference during human-computer interaction have targeted the physiological/motivational state of valence - the hedonic dimension of emotional states as being either positive or negative. Much like camera recordings, electromyography (EMG) has been applied to detect facial expressions indicating positive or negative affect \cite{barral2016extracting} and emotional expressions, even in response to unconscious primes \cite{dimberg2000unconscious}. Electrodermal activity is measured by recording changes in the conductance of the skin, as affected by the sympathetic nervous system's effect on sweat glands, and thus used to indicate arousal. Classically, EDA has mainly enabled the detection of arousal during human-computer interaction \cite{herbon2005emotion}. More recently, advanced neural network models have shown EDA can also be used to accurately infer valence \cite{yu2020systematic}. Measures derived from electrocardiography, such as the orienting response (a brief slowing of the heart rate in response to stimulus) and heart-rate variability, have been proposed to index both arousal \cite{bonnet1997heart} and valence \cite{gruber2015heart}, and are commonly used indices to index user experience, preference, and evaluation. 

Beyond electrophysiology of the muscles, skin, and heart, preferences can be gauged using measures that are commonly seen as more \say{directly} related to affect and cognition. Such methodologies include functional magnetic resonance imaging (fMRI), fNIRS, and EEG, which quantify activity within the brain \cite{spape2015human}. Functional magnetic resonance imagery (fMRI) measures the blood-oxygenation-level-dependent (BOLD) signal. Neural tissues draw on oxygen and other nutrients to sustain their activity, allowing precise spatial mapping of brain activity. While fMRI remains a prohibitively costly tool used mainly to study brain function, it has also been used in HCI contexts, such as in detecting relevance \cite{moshfeghi2013understanding}. 

Previous work on measuring neurophysiology for user modeling has mainly focused on correlating physiological states to specific cognitive functions: emotional or affective responses \cite{white2017improving, Barral:2015:EPP:2678025.2701389,Kim2004}, cognitive load \cite{Hincks:2016:UFR:2977034.2977055}, and relevance \cite{Eugster:2014:PTB:2600428.2609594,Jacucci:2019,BlaAcqDaeHauSchStuUscWenCurMue16,Arapakis:2008:AFI:1390334.1390403}. The computer science community has been focusing on understanding how information needs arise \cite{Moshfeghi:2016,Moshfeghi:2013UNDERS} and how relevance or binary preferences are manifested in EEG \cite{Kangassalo:2019:WUI:3331184.3331243,Wenzel:2017}, MEG \cite{Kauppi2015288} and fMRI data \cite{Moshfeghi:2013}. Recent research has also utilized these signals to predict relevance \cite{Eugster:2014:PTB:2600428.2609594} and use it as a feedback mechanism to enable interactive recommendation of information directly from brain measurements \cite{EugsterSREP}. 

Event-related potentials (ERPs), brain responses evoked by a variety of mental and physical events, can provide valuable information on user processing by indexing attention \cite{polich_updating_2007}, task relevance, and informativeness \cite{sutton_information_1967}.  Most applications using ERPs rely on an explicit type of processing (e.g. BCI spellers, \cite{FARWELL1988510}), but ERPs have also been targeted in measuring the degree to which a user has implicit information \cite{rosenfeld2005brain}. For example, EEG variants of the guilty knowledge test aim to detect whether suspects recognize presented information concerning a crime scene that an innocent person could not possess, despite giving explicit information to the contrary \cite{rosenfeld2005brain, farwell2012brain, meijer2013comment}. Similarly, distinct ERPs have also been shown to reveal when implicit memories are activated, even when users are not explicitly aware of having such memories \cite{rugg1998dissociation}. Therefore, EEG measurements are likely to contain information that can reveal preferences without any physical interaction with a computing system,  even in cases where users are themselves not aware of their preferences, or are prone to intentionally express false preferences. 

Despite these achievements, the majority of previous work is focused on relevance (i.e. binary measure of whether the information is relevant or not to a given topic) and does not always reflect the more complex phenomena of subjective preferences. Furthermore, the accuracy levels reported in previous research \cite{Eugster:2014:PTB:2600428.2609594, Kauppi2015288, Arapakis:2008:AFI:1390334.1390403}, while demonstrating the feasibility of neurophysiological signals in relevance detection, are still not at the level of pragmatic use in learning accurate representations of user preferences and needs. Finally, the added value of using these signals is unclear; if neurophysiological signals bring in more information about users than behavioral signals alone, are they also more reliable? 

The promise of physiological signals for inferring preferences is twofold. First, the human cognitive system can be monitored continuously and can therefore provide high throughput data that only requires the user to attend to particular digital information. That is, users are not required to explicitly interact with the content, but measurements can be done whenever users \textit{perceive} the content \cite{Eugster:2014:PTB:2600428.2609594}. Second, user preferences inferred directly from the human cognitive system could complement potentially inaccurate inferences from other behavioral signals.

\section{Neurophysiological data acquisition}

In this section, we describe an experiment where electroencephalography (EEG) data of thirty-one participants were recorded in response to viewing images of faces. The participants' corresponding self-reported graded preferences for these images were also obtained. We then explain the featurization procedure that was applied to these data to study the relationship between brain responses and explicit graded preferences, as well as investigate if and when brain responses diverge from explicit preferences.

\subsection{Participants}
Thirty-one volunteers were recruited from the University of Helsinki. The participants were fully informed of the nature of the study and signed informed consent to acknowledge understanding of their rights as participants in accordance with the Declaration of Helsinki, including the right to withdraw at any time for any reason without fear of negative consequences. Complete data were obtained for 31 participants (18 male and 13 female). The mean reported age was 28 years (SD = 7.14, range 18-45). For gender preferences, 18 participants reported that they preferred females, 8 preferred males, and 5 reported equal preference between males and females (recorded via an on-screen prompt). All participants were compensated for their participation with two cinema vouchers. 

\subsection{Apparatus}\label{sec:apparatus}
An LCD monitor, positioned 60 cm from the participants with a resolution of 1680 by 1050 pixels and running at 60 Hz, was used to present stimuli.  The timing of display and EEG amplifier trigger control were optimized using Psychology Software Tools E-Prime 3.0.3.60 stimulus presentation software. EEG data were recorded using 32 Ag/AgCl electrodes, positioned at equidistant locations of the 10-20 system using EasyCap elastic caps (EasyCap GmbH, Herrsching, Germany). A QuickAmp USB (BrainProducts GmbH, Gilching, Germany)  amplifier running at 2000 Hz was used for hardware amplification, filtering, and digitization of the data. Two pairs of bipolar electrodes placed 1 cm lateral to the outer canthi of the left and right eyes, and 2 cm inferior and superior to the right pupil, were used to collect the electrooculogram (EOG). Data collected from the EOG were used for artifact removal.

\subsection{Stimuli}

An important aspect of the experiment was to use stimuli that would be relatively easy for the participants to assess their subjective preferences. Therefore, the experiment was designed as a human face image attractiveness experiment. Using faces as stimuli allows capturing the preference reactions of the participants, as humans have a natural rapid preference reaction in response to seeing a face \cite{Luo2019Robust}, and this reaction may naturally vary between high preference and low preference \cite{olson2005facial}.

It was also required that the participants not recognize the stimuli so that their personal assessment of the stimuli would be based on perceived preference rather than a participant's confounding pre-knowledge of the content appearing in the image. For example, using publicly available images could introduce additional confounding factors that can not be easily accounted for. Evoked responses from participants may be produced from viewing a face they simply recognize, but do not necessarily find personally attractive. Additionally, real-world stimuli often contain a high degree of internal variation which could produce neurophysiological effects unrelated to preference assessment. To avoid this, we decided to use images of artificially generated faces as stimuli. This allowed us to use relatively homogeneous stimuli (that is, with a similar pose and orientation) that did not represent any known individual. 
Stimuli were produced using a generative adversarial network (GAN) trained on a large dataset of celebrity faces \footnote{http://mmlab.ie.cuhk.edu.hk/projects/CelebA.html}. Images were generated by applying a parameter randomization over the 512 vectors by sampling via a random process from 70,000 latent vectors from a 512-dimensional multivariate normal distribution \cite{karras2017progressive}. Each generated face was manually screened by a human assessor to ensure there were no errors in the generated images, such as having zippers for mouths or other unrealistic distortions, but without regard for any other physical attributes. The images were then sorted into male- and female-appearing groups, of which the first 120 images were selected, for a total of 240 unique images. These images were compressed in resolution to 512 x 512 pixels, and an elliptic grey frame was applied to mask background and non-facial features. 

\subsection{Procedure}
The experiment consisted of 3 blocks; for each block, all 240 images were presented twice in 8 rapid serial visual presentation (RSVP) trials of 60 images each and a confirmation task following each trial (see Figure \ref{fig:experimental_protocol}). For each participant and for each trial, images were presented in random order. Participants were instructed to simply observe the presented faces and count the faces they found personally attractive. To enhance task salience, 4 random images (either male or female, randomized) were provided at the beginning of each trial, and participants were asked to click on the one they found least attractive (these images were not used as stimuli in any of the RSVP trials). The RSVP then started, with 60 images presented at 2 Hz (500 ms stimulus onset asynchrony) against a grey background. At the end of each RSVP trial, participants were asked to report the number of images they had counted as personally attractive. Participants were then presented with a confirmation task, where smaller versions of the 60 images were provided. The participants were requested to select all of the images that they had found personally attractive, with their self-reported count being displayed as a reminder. The experiment took ca. 25 minutes, excluding two self-timed breaks between the three blocks. 

\subsection{Data preprocessing}
The recorded EEG data were digitized and online referenced to the common average. Data preprocessing included a band-pass filter applied between 0.2 and 35 Hz to remove slow signal fluctuations and line noise. Then, the data were time-locked to stimulus-onset and split into 1100 ms long segments (henceforth referred to as ``epochs'') including 200 ms of baseline activity. For each epoch, the average of its baseline activity was subtracted from the remaining 900 ms of activity. Finally, epochs containing artifacts (such as eye blinks) were excluded from the data set by applying an individually tailored thresholding to the absolute maximum voltage. This resulted in the removal of approximately 12.2 percent of all epochs, with the final dataset comprising 1265 (SD = 109) epochs per participant, out of a maximum possible 1440. 

\subsection{Data labeling of explicit and group preferences}

For the explicit preferences experiment, we used all epochs from the post-processed data. Each epoch was associated with a unique image ID, participant ID, and participant-specific explicit preference rating (an integer from 0 to 3, based on the number of times the image was selected across RSVP tasks for that participant).

\begin{figure*}[!ht]
    \centering
    \includegraphics[width=\textwidth]{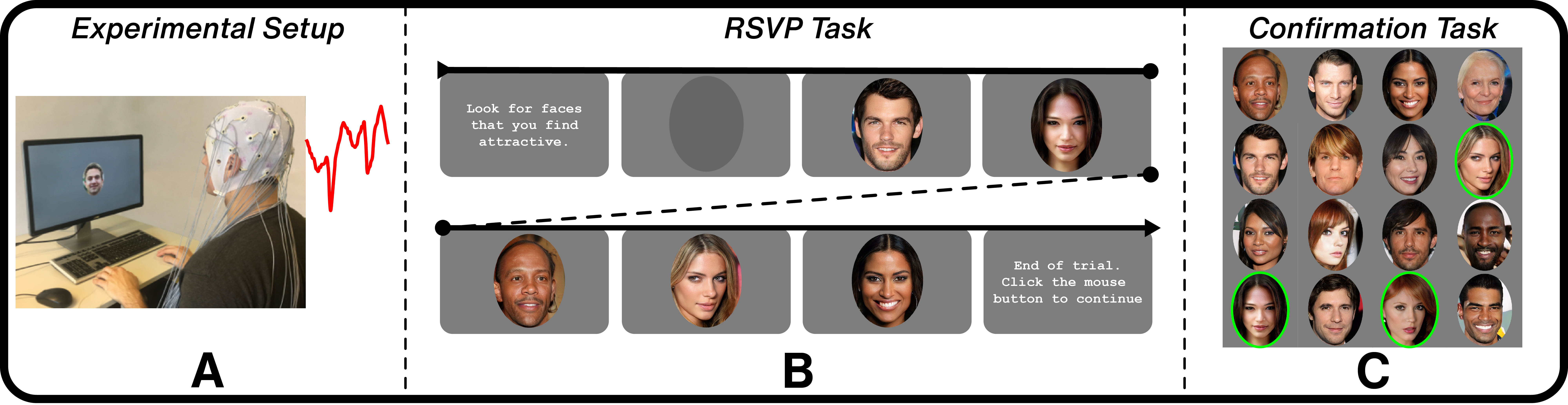}
    \caption{Diagram of the experimental procedure. A). A subject is fitted with an EEG cap, situated in front of a computer screen presenting stimuli while brain responses are recorded. B). Detail of the RSVP task, where a subject is prompted to look for images of people they find personally attractive while the images are presented in random order every 500ms in rapid-serial visual presentation format. C). After viewing a batch of images, the subject is then asked to confirm which images depict a face they found personally attractive during the RSVP task. }
    \label{fig:experimental_protocol}
\end{figure*}

For the group preferences experiment, the group preference ratings were assigned using a popularity model, calculated by the rank-ordered sum of all individual's self-reported ratings of presented images. For the group preferences experiment, we selected from our EEG data only the epochs associated with stimuli that individuals did \textit{not} self-report as attractive (i.e. stimuli were never selected during the confirmation tasks). These data were then separated into two distinct categories: epochs for stimuli in which the group preference placed the stimuli in the top third, and epochs for stimuli in which the group preference placed in the bottom third. 

\section{Explicit preference inference}

The purpose of the explicit preference inference experiment was to study whether an individual's subjective self-reported preferences could be associated with ERPs and classified in a single-trial machine learning scenario. 

\subsection{Neurophysiological findings}

\begin{figure*}[!ht]
\centering
  \includegraphics[width=0.40\textwidth]{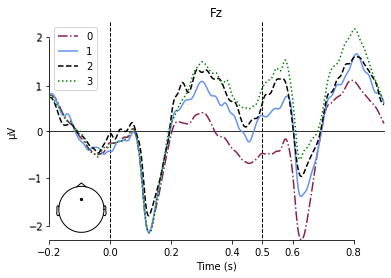}
   \includegraphics[width=0.40\textwidth]{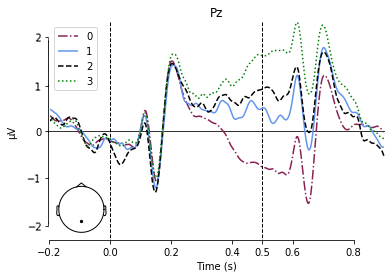}
  \caption{Grand average voltages of each ERP-component, grouped by the number of times each individual selected a stimulus during the confirmation tasks, at the Fz (left) and Pz (right) channels. Stimuli onsets are indicated by vertical dashed lines at 0.0 (s) and 0.5 (s). The effect is shown as a strong scalp positivity, particularly at the Pz site, which increases the more an individual preferred a stimulus. A small scalp negativity effect appears for stimuli that an individual did not select. }~\label{fig:erp_plot_explicit}
\end{figure*}

Epochs for stimuli were grouped by their calculated explicit preference rating (the number of times an individual participant selected a stimulus across confirmation tasks). Then, they were averaged across all participants and analyzed by channel to construct ERP plots. Visual inspection of the grand average ERPs suggested large differences between preference ratings over frontocentral (Fz, Cz) and temporoparietal (TP9, TP10, P7, P8, and Pz) sites. Given the literature on functional and topographical differences indicated by a frontal P3a and parietal P3b \cite{polich_updating_2007}, we further analyzed the effect of preference on the evoked potential observed for the Fz (frontal) and Pz (parietal) electrodes, as shown in Figure \ref{fig:erp_plot_explicit}. To investigate the time course of the effect, we used a series of repeated measures ANOVAs with a sliding window, averaging the data in bins of 25 ms, over the stimulus-evoked amplitude from 100 - 500 ms. No significant differences were found between preference levels of \textit{1} and \textit{2}, so they were combined for the remainder of the neurophysiological analysis. Repeated measures ANOVA showed for Fz a significant effect from 250 ms, F (2, 60) = 13.15, p = .0003, reaching a maximum effect at 350 ms, F (2, 60) = 28.54, $p < .0001$, and remaining significant between 250 and 500 ms, \textit{F}s $> 13$, \textit{p}s $< .0004$. For Pz, a similar effect was observed slightly later, starting at ca. 275 ms, F (2, 60) = 7.14, p = .03, but unlike Fz, increasing in size and being maximal at 500 ms, F (2, 60) = 68.47, $p < .0001$, with every point in between being significant also, \textit{F}s $> 7.24$, \textit{p}s $< .03$. Post-hoc pairwise comparisons at the time of maximal effect for Fz showed a significant difference between \textit{1 \& 2} and \textit{0}, T(30) = 6.99, $p < .0001$, and between \textit{3} and \textit{0}, T(30) = 6.54, $p < .0001$, but not between \textit{3} and \textit{1 \& 2}, T(30) = 1.69, p = .10. Conversely, for Pz, there was a significant difference between preference ratings, \textit{T}s $> 4.9$, \textit{p}s $< .0001$. In other words, graded preference was found to occur at Pz, while Fz indexed a more binary type of preference. 

\subsection{Classification setup and feature engineering}

\subsubsection*{Classification setup} 
We used the same explicit preference EEG data used in the neurophysiological analysis, grouped by individual participant rather than by rating category. 

\subsubsection*{Feature engineering} Vectorized representations of the EEG data were constructed \cite{blankertz2011single} by splitting each epoch into 15 equidistant time windows on the 50 - 800 ms post-stimulus period, and then averaging the EEG measurements in each window. This resulted in data tensor $X^{n \times m \cdot t'}$ for each participant. Here, $n$ denotes epochs, $m = 32$ channels and $t' = 15$ the time windows. Each epoch $n_i$ also had an associated preference rating $Y_i$. Respectively these are the inputs and outputs of the classifier models.

\subsection{Model and evaluation}
\subsubsection*{Model} Individual preferences were decoded in a single-trial ERP classification scenario using methods similar to those presented in \cite{blankertz2011single}. A regularized Linear Discriminant Analysis (LDA) classifier was used, with shrinkage automatically selected using the Ledoit-Wolf lemma \cite{ledoit_well-conditioned_2004}. For each participant, two classifiers (one for explicit preferences and one for implicit) were trained and evaluated using a leave-one-out split of each participant's corresponding vectorized EEG data.

\subsubsection*{Evaluation} The classifier performance was measured using the Area Under the ROC Curve (AUC) measure. A baseline was constructed by using a classifier trained with randomly permutated class labels \cite{ojala2010permutation}. This allowed computing permutation-based ($n$=$1000$) p-values acquired by comparing the AUC scores of the random classifiers to those of classifiers trained for each participant.

\subsection{Classification results}
\begin{figure}[!ht]
\centering
  \includegraphics[width=1\columnwidth]{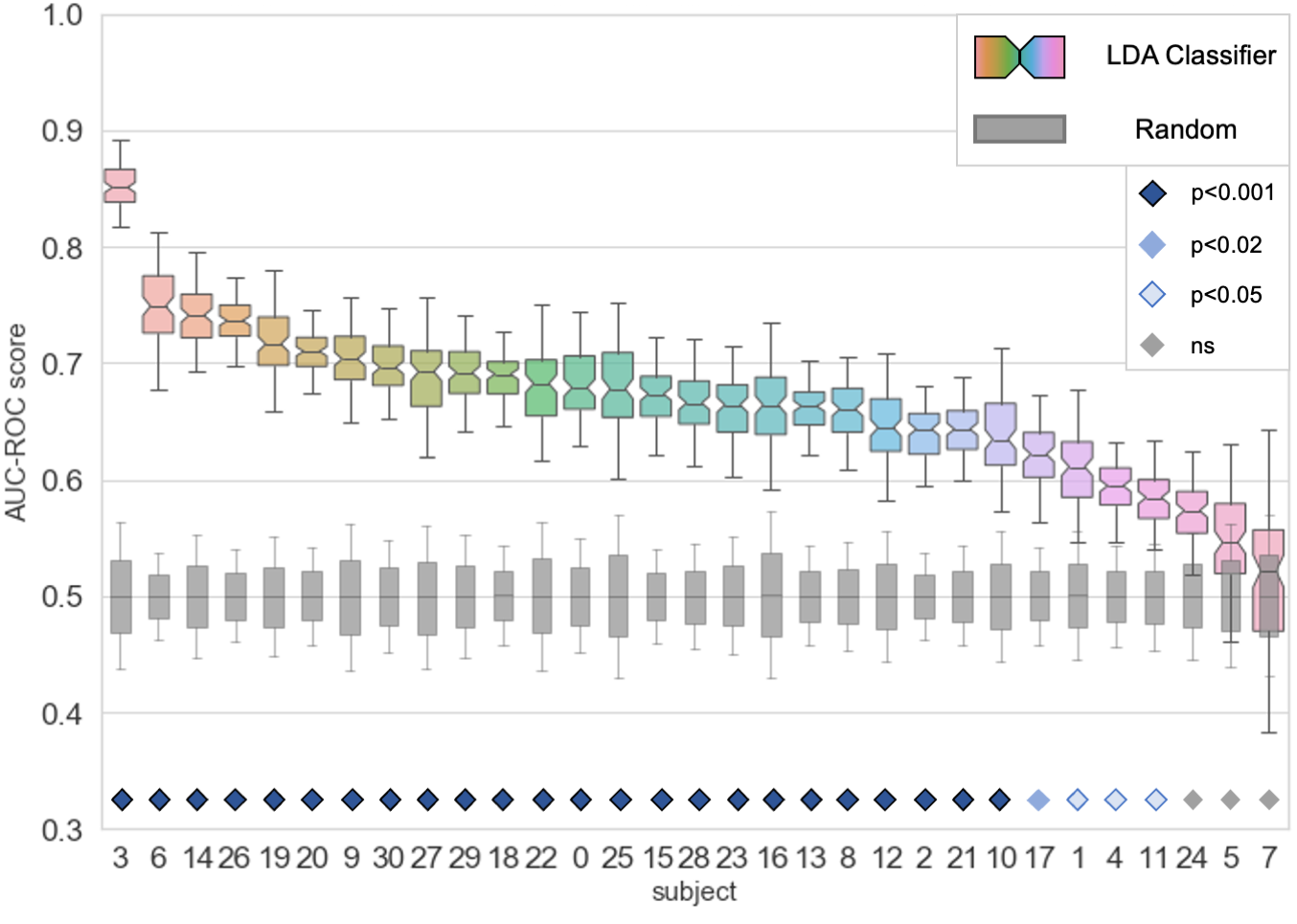}
  \caption{AUC scores for the per-participant explicit preference classifiers, ordered by median AUC score. The classifiers successfully predicted the four levels (0, 1, 2, and 3) of preference for 28/31 participants with a mean AUC = 0.66.} ~\label{fig:boxplot_ynm}
\end{figure}

\begin{figure*}
\centering
  \includegraphics[width=0.75\textwidth]{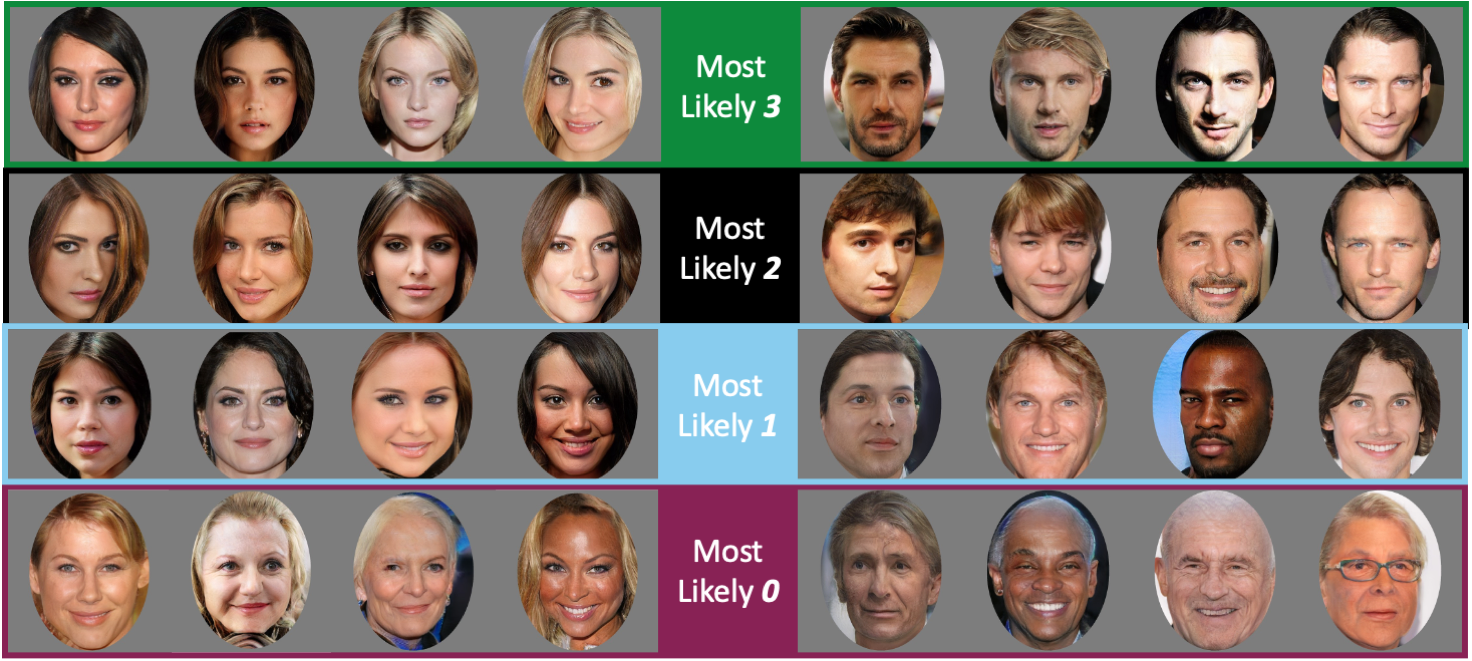}
  \caption{Representative sample of stimuli images which returned the highest log probabilities for the different preference levels in the explicit preferences experiment, averaged across all per-subject classification models. Stimuli are grouped into female-preferring and male-preferring cohorts.} ~\label{fig:exp_pref}
\end{figure*}

Explicit preferences were successfully classified for an overwhelming majority of participants. The LDA classifiers performed better than random for 28 out of 31 participants, with a mean AUC score of 0.70 and a standard deviation of 0.08 across all participants. Participant-level AUC scores and significance levels are shown in Figure \ref{fig:boxplot_ynm}. Representative samples for classifier outputs are provided in Figure \ref{fig:exp_pref}. A Wilcoxon signed-rank test revealed ($p < .0001$) that classifiers for female-preferring participants were significantly more accurate (AUC = 0.70) than for male-preferring participants (AUC = 0.64), which aligns with the existing psychological literature \cite{wood2009using}.

\section{Group preference inference}

The purpose of the group preference experiment was to study if it is possible to uncover additional preference information beyond what was explicitly reported by the participants and that manifested in EEG activity.

\subsection{Neurophysiological findings}

EEG data was subset from the entire dataset by averaging grouping the epochs according to their group ranking (third tertile = High, second tertile = Medium, first tertile = Low). 

\begin{figure*}[!ht]
\centering
  \includegraphics[width=0.40\textwidth]{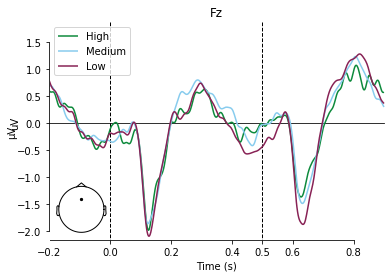}
   \includegraphics[width=0.40\textwidth]{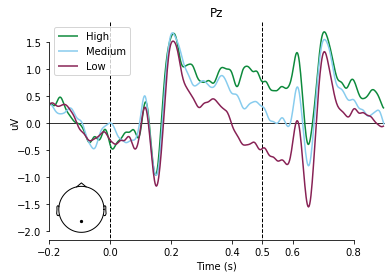}
  \caption{Grand average voltages of each ERP-component for the group preferences experiment. stimuli that subjects did not consider attractive according to their self-reported ratings, grouped by the tertile of the group preference, at the Fz (left) and Pz (right) channels. Stimuli onset are indicated by vertical dashed lines at 0.0 (s) and 0.5 (s).} ~\label{fig:erp_plot_implicit}
\end{figure*}

Again, we analyzed the channels most likely to reflect P3a and P3b potentials by conducting a sliding series of repeated measures ANOVAs for the temporal range between 100 and 500 ms with the average evoked amplitude within 25 ms bins as a measure. As can be clearly seen in Figure \ref{fig:erp_plot_implicit}, this type of preference showed no effect on Fz, Fs $< 3.8$, ps (Bonferroni adjusted) $> .5$, in any bin. However, for Pz, an effect was observed at 275 ms, F (2, 48) = 8.52, $p < .01$, and from 350 ms onwards, Fs $> 6.79$, \textit{p}s $< .05$, with a maximal effect at 500 ms, F (2, 48) = 18.03, $p < .0001$. Post-hoc comparisons on this bin between the three preference levels showed a significant difference between \textit{Group Medium} and \textit{Group Low}, T (24) = 5.99, $p < .0001$, and between \textit{Group High} and \textit{Group Low}, T (24) = 5.29, $p < .0001$, but not between \textit{Group High} and \textit{Group Medium}, T (24) = 1.92, $p < .07$. Thus, a dissociable effect of explicitness of preference could be discerned on the electrode level: Pz amplitude showed a clear effect of popular preference level even in the absence of individual, explicit rating, while Fz was unaffected. This finding suggests that while participants may not express a preference, their brain responses nonetheless correlate with peer group preferences. 

\subsection{Classification setup and feature engineering}

\subsubsection*{Classification setup} 
The EEG data used for the group preference experiment was the same data used in the neurophysiological experiment and obtained via the procedure from Figure \ref{fig:experimental_protocol}. Because the neurophysiological findings found no significant difference between EEG signals in the top and middle tertiles, epochs for stimuli ranked in the middle tertile were omitted from the analysis.

\subsubsection*{Feature engineering} 
We used the same features and classifier evaluation strategy as in the explicit preference inference experiment, using the group preference labels as the classification targets.

\subsection{Classification results}
\begin{figure}
\centering
  \includegraphics[width=1\columnwidth]{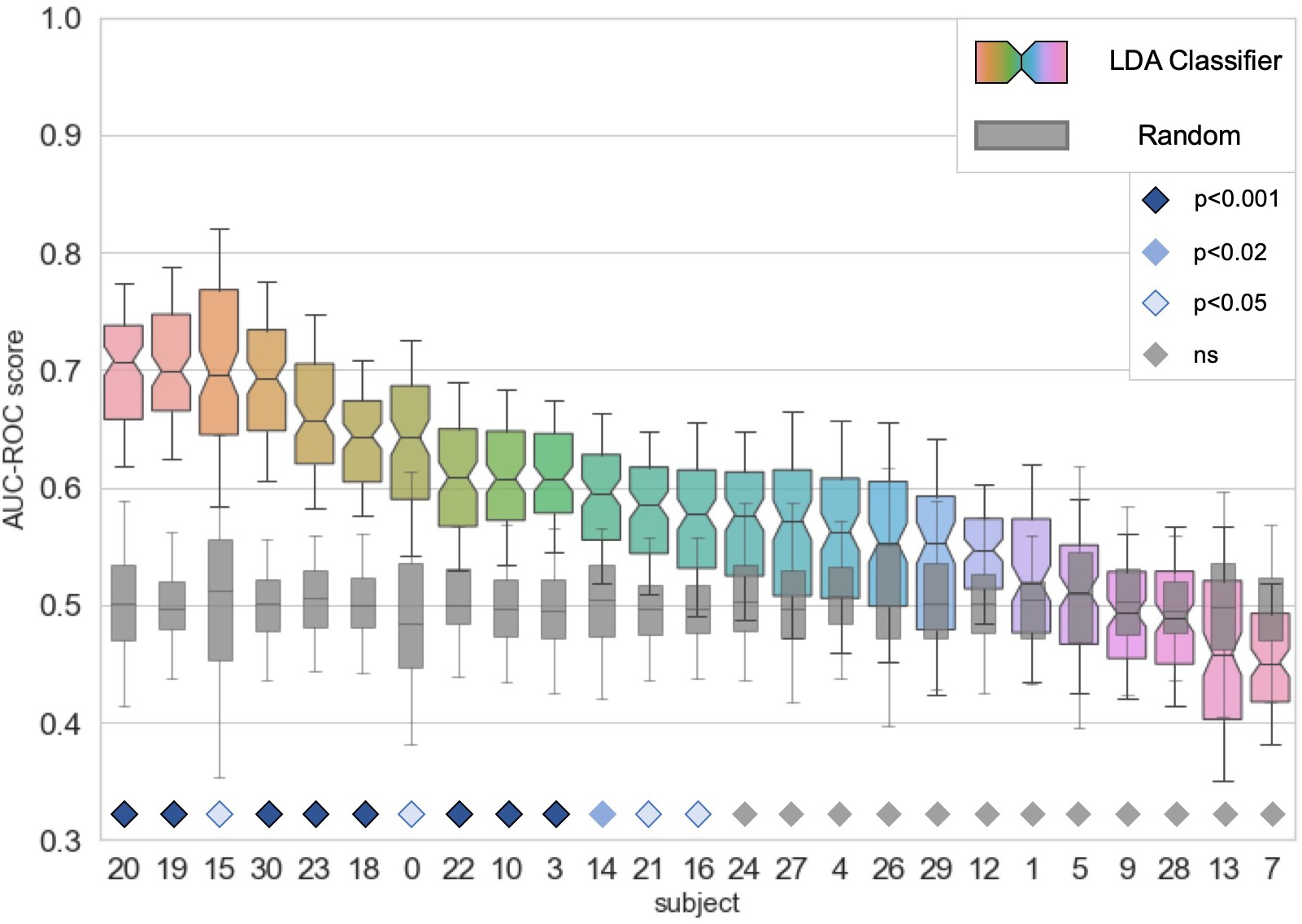}
  \caption{AUC scores for the per-participant group preference classifiers, ordered by median AUC score. The classifiers successfully predicted the group preferences for 13 out of 25 participants. 6 participants could not be included in the analysis, as their self-reported preferences matched the group preference.} ~\label{fig:boxplot_preffalse}
\end{figure}

\begin{figure*}[t!]
\centering
\includegraphics[width=0.95\textwidth]{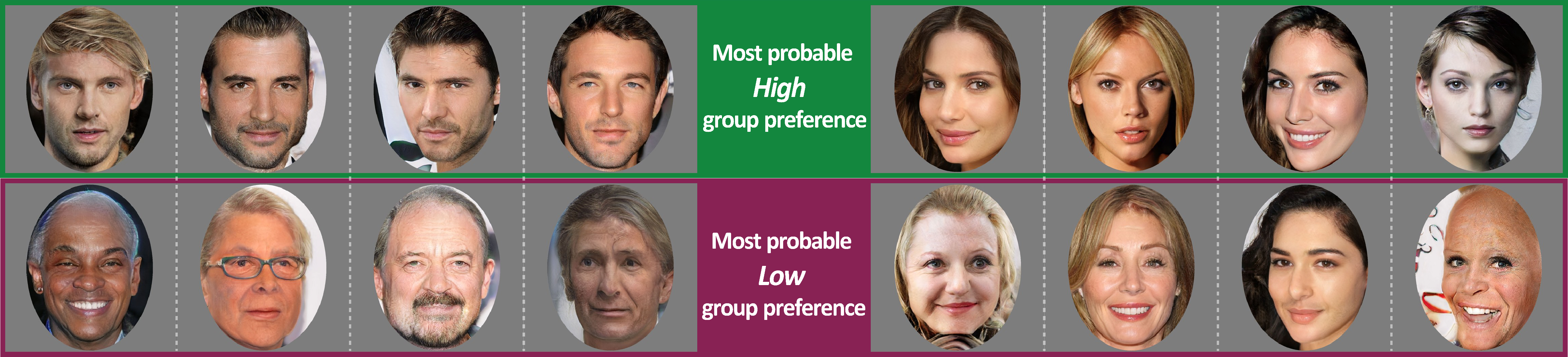}
\caption{Representative sample of stimuli images for which the associated EEG-responses returned the highest log probabilities for the "High group preference" or "Low group preference" classes. These were sampled from the results of 8 randomly chosen participants (4 female-preferring and 4 male-preferring).} ~\label{fig:imp_pref}
\end{figure*}

Classification could not be performed for 6 of the 31 participants, either because they had marked all presented stimuli as ``attractive'' at least once, or because their explicit preferences directly reflected the group preference, i.e. all stimuli that had the \emph{explicit} label ``No'' were also in the bottom tertile of group preference. For the remaining 25 participants, preferences were successfully classified for 13 of them, where the LDA classifiers performed better than random, with a mean AUC score of 0.66 (SD = 0.10). Across these 25 participants, classifiers achieved a mean AUC score of 0.60 (SD = 0.13). Participant-level AUC scores and significance levels are shown in Figure \ref{fig:boxplot_preffalse}, and representative samples of classifier outputs are shown in Figure \ref{fig:imp_pref}. A Wilcoxon signed-rank test revealed ($p < .0001$) that classifiers for male-preferring participants were significantly more accurate (AUC = 0.62) than classifiers for female-preferring participants (AUC = 0.58) (the opposite effect seen from the explicit preferences experiment). 

\section{Discussion and Conclusions}
Despite advances in the past few decades in recommender systems and content personalization, most of the signals used to estimate the preferences of users are captured from behavioral measurements that are assumed to be indicators of user preferences. These signals have been found to be reliable probes for the users' underlying preferences in various studies, but evidence also shows that these predictions are not always satisfactory for the users. Researchers have expressed concerns about the reliability of using behavioral signals as evidence of the methodological performance of preference prediction \cite{Knijnenburg:2012}. 

For example, a typical dataset of movie recommendations contains click and rating data. These data are used both for training the models and are also often used for evaluating the classifiers. Yet, we know fairly little about whether the ratings reflect the real preferences experienced by the users, and how much the results generalize outside of the evaluation domain and application from which the data was collected from. Rather than using indirect and potentially confounded behavioral cues to estimate preferences, we argue that recording physiological signals provide more direct estimations of individual preference. Here, we have demonstrated the feasibility of a new method of preference estimation that relies solely on the brain responses of an individual. We also have demonstrated that brain responses contain additional information that can directly contradict an individual's explicitly stated preferences for half of our experimental participants. This phenomenon warrants further investigation, as its origins are not entirely clear.

\subsection{Summary of contributions} 

We studied the association between a user's self-reported, graded preferences and their evoked brain activity. We also studied whether or not evoked brain activity could reveal additional preference information. EEG recordings measuring responses to images in a facial attractiveness task were used to test whether individual and group preferences could be inferred from brain signals. To investigate whether preferences could be inferred from brain signals and if they provide accurate information about the underlying real preferences of users, we asked two research questions that we answer below: 

\textbf{RQ1}: Can brain responses be used to estimate graded user preferences?

\textbf{Answer}: Our results show that brain responses can be reliably associated with self-reported preferences, and these responses are graded according to preference level. Results from both the neurophysiological and machine learning experiments show statistically significant effects for personal explicit preferences. Furthermore, the machine learning experiments demonstrate that it is possible to classify explicit graded preferences for unobserved responses in a single-trial setting.  

\textbf{RQ2}: Can group preferences be estimated from an individual's brain responses?

\textbf{Answer}: Our results show that brain responses often, but not always, correlate with group preferences. Furthermore, we find that brain responses correlate with group opinion even when an individual explicitly states no such personal preference exists, suggesting brain responses can contradict an individual's explicitly stated preferences. The effects were dissociable in the topography of ERPs in the parietal area. We also demonstrate a machine learning experiment that successfully classified high and low group preferences from an individual's brain responses when the individuals had explicitly stated no such preference.

In addition to achieving reasonable classification performance, the results are further validated by visual inspection of the representative outputs shown in Figures \ref{fig:exp_pref} and \ref{fig:imp_pref}. These results seem to show an accurate reflection of the general aesthetic preference of the participant population. As is immediately obvious from Figures \ref{fig:exp_pref} and \ref{fig:imp_pref}, stimuli that appear older than the median participant age (28) were less likely to be classified as highly preferred by the models. This is consistent with previous observations suggesting people generally prefer younger adults over older ones when selecting for sexual attraction \cite{silverthorne2000sexual}. 

Critically, we demonstrated that EEG can capture information that enables us to infer additional preference information beyond what is explicitly stated. In particular, we show that faces that are rated as attractive by most people, but not a specific participant, produce EEG responses indicating the participant distinguishes between this popularly attractive face and others, despite the participant explicitly denying any attraction. This shows EEG can be used to detect preference information beyond what is available via explicit self-report. 

\subsection{Limitations and future work}

Our study was based on a stimuli set comprised of artificially generated facial images rather than content typically used in personalization research, such as songs, films, and news articles. While this does limit the extent to which we can confidently generalize our findings towards other types of content, related work such as \cite{pinkosova2020cortical} has already identified a similar graded relevance phenomenon using text as stimuli, and there is a growing body of literature across imaging methods that document graded brain responses for a variety of phenomena \cite{scheerer2013erp, kangassalo2020information}. Thus, we suspect that our findings will generalize to other domains. Furthermore, the use of face images provides significant advantages over using other real-world media data for this foundational work. 

First, visual assessments of facial attractiveness are naturally and rapidly made by humans. As our experiments rely upon EEG data collected through RSVP trials, it requires stimuli that can be quickly assessed by an average person. Movie posters or more visually complex images as stimuli would require participants to spend significantly more time evaluating each stimulus and would violate the RSVP paradigm.

Second, using artificially generated faces ensures all participants are being exposed to stimuli that they have never seen before, and of similar visual features (oval cutout of a face on a grey background). In this way, data is not confounded with stimuli variation that could cause effects that are not related to preferences. For example, video content or more general image content would have greater natural variation in color spectrum and content density, and video would additionally have variance in terms of editing style or rhythm. Such confounds could cause the brain responses to reflect those stimuli features rather than the preference effect.

Our approach also largely relies upon the ability to identify and distinguish between time-locked ERPs in response to still imagery. Using stimuli that vary across time, such as musical arrangements, in addition to presenting unique variance challenges described above, would also require an approach that can capture longer-lasting effects or models that can target particular temporal windows. Classifying preferences for stimuli that vary significantly in their attributes across time is beyond the scope of the present study.

\subsubsection*{Group Ranking Approach}
To detect group preferences, we relied upon featurizing our dataset based on the collective self-reported responses of all participants. This assumes that the participants, split by gender preferences, share more or less the same ideals of attractiveness. In this experiment, the relative homogeneity of the participant pool - university students of predominantly European descent, with some participants of East or South Asian descent - allowed for such an approach. With a much larger and more diverse set of participants, where there may be multiple and different cultural perceptions of beauty, it may not be sufficient to use the global ranking of stimuli across all participants, and instead necessary to first cluster participants by cultural background. However, this can introduce its own set of biases into the results and must be done with care.

\subsubsection*{Neurophysiological Findings}

It is not entirely apparent which cognitive effect is found when brain responses correlate with group preference and contradict an individual's explicitly stated preferences. Previous work has investigated the neural markers of \textit{consciously} violating \textit{group preference} \cite{trautmann2013eeg}, however, such work does not generalize to participants reporting their preferences without any conscious knowledge of how others responded. It could be that individuals indeed found faces attractive but consciously chose to \textit{not} select those faces during the explicit confirmation task due to encoded social pressures or bias (``this face is attractive, but not my type'').  Alternatively, they were not consciously aware of whether or not they found the face attractive, and instead, some other phenomenon is being recorded, such as a divergence of personal preference from encoded knowledge of group-wide preferences. This divergence was predicted better than random in about half of the participants for which data was available (13 out of 25), and less than half considering all participants (13 out of 31). Nonetheless, more than a third of participants displayed some sort of discrepancy in their explicit choices that was detectable in the brain, suggesting that additional information was captured from brain signals that is otherwise missing from explicit preference judgments. Independent of the exact cause of this effect, which requires further investigation, our data reveal a strong association between group preferences and an individual's brain responses, which can be detected using both neurophysiological markers as well as through machine learning.

Even with the limitations described above, we believe that our approach constitutes an important step in revealing and predicting graded preferences directly from brain signals. While our experiments are conducted within a subjective facial attractiveness context, we believe that the effect is not restricted to the present experiment and can be extended to a variety of preference elicitation and personalization scenarios.

\subsection{Ethical implications}
While widespread adoption of brain-computer interfacing as a primary mode of interaction is unlikely to occur within the next decade, as the technologies associated with it improve and the necessary equipment becomes less expensive and more user-friendly, it is reasonable to expect that the use of BCIs will become more common in the coming years. Brain-monitoring applications have already been introduced by state organizations and private companies in China with the intention of boosting worker productivity and improving the learning outcomes of students \cite{chan_2018}. Such applications, however, have been met with skepticism and caution, and there is certainly reason to be concerned regarding the early adoption of this technology.

Consider how complex user modeling to date has mostly been feasible only in "big data" contexts. Consequently, most public concern regarding user modeling and data collection has been directed toward organizations with access to millions of user profiles through proprietary apps or platforms where data on many implicit interactions may be collected. As the use of physiological signals as an input modality becomes more common, advanced behavioral modeling of individual users may no longer require access to large numbers of user profiles or extensive amounts of recorded explicit and implicit user interactions. BCIs could therefore enable small organizations with access to only a few users to build complex models of user behavior using fairly limited amounts of data. Indeed, an emerging field known as \textit{neuromarketing} utilizes neurophysiological information to improve the success of targeted advertisements and other marketing strategies \cite{khurana2021survey}.

Although legislation such as General Data Protection Regulation (GDPR) introduced in the European Union may offer some protection for the average person, it does not solve the problem of individuals consenting to data collection when neither they nor the data collectors fully understand the extent to which such data could be used in the future. Consider the findings of this work, which demonstrate evidence of additional preference information detectable via EEG. The implications of discovering preferences outside of a user's awareness and without their knowledge or consent are enormous, as a user may have only consented to data collection that they had assumed could not reveal such information.%It should be assumed that the general public's understanding of BCI capabilities will always lag behind what is \textit{actually} possible, which calls into question their ability to Terms of Service %even Furthermore, it is to be expected that private companies will not publicly disclose discoveries they have made regarding novel ways in which the data they have collected can be exploited to reveal more information about a given user, particularly if their Terms of Service and User Agreements have been written as sufficiently vague to include such novel discoveries.

Even with a well-informed general public, the use of BCIs carries risks that are unique from existing modes of interaction, as it involves recording data that could be used to reveal medical conditions. EEG equipment has long been used to detect neurological disorders such as epilepsy, while more recently fNIRS has demonstrated promise in the diagnosing of Alzheimer's Disease \cite{perpetuini2019complexity}. Additionally, since raw data collected from an individual via EEG and other BCI methods can be used to identify them between platforms with relative ease \cite{chakladar2021multimodal}, ubiquitous adoption of BCIs poses significant risks if the associated BCI data is improperly managed and anonymized. Future security breaches leaking BCI data could affect an individual across multiple services and platforms.

\subsection{Implications for human-computer interaction}

In essence, users' preference decisions toward digital content happen in the brain in response to perceiving content. So far, our understanding as to whether the preferences are accurately captured by behavioral cues, or even when explicitly expressed by users, has been limited. That is, previous studies have relied on methods based on either directly requesting input about user preferences or observing users' ratings or interaction behavior with content delivery systems to infer subjective preferences. Preferences estimated from interaction, such as from self-reported ratings or implicit behavioral patterns, are at best only probes for underlying, subjective preferences. Implicit behavioral signals tend to be uncertain as users may click  \cite{Joachims:2002:OSE:775047.775067} and view content \cite{Buscher:2012:ADE:2070719.2070722} even if they do not prefer the content, for example to, develop a better idea of what the content is like.

Our research is the first to show evidence that graded user preferences toward digital content have direct correlates with brain potentials evoked in response to perceiving the content, and that preferences can be predicted in a single-trial scenario from brain signals. By using artificially generated images of faces as stimuli, participants were not influenced by prior attitudes toward the content, suggesting that graded preference judgments have a natural origin that should generalize to new information and different preference elicitation scenarios. Our results also revealed additional preference information conflicting with self-reported ratings, e.g. preferences that do not match the explicitly reported preferences by users but which do correspond to the group preference, could be revealed and predicted from brain signals. This suggests that brain signals can capture preference information that cannot be revealed through explicit interactions. These results challenge the simplistic assumptions made about the reliability and representativeness of self-reported ratings and behavioral data of user preferences used as \emph{de facto} benchmarks in a variety of research, including the development of recommendation and personalization methods.

We believe our findings can inform both the human-computer interaction community and personalization system researchers on the potentially limiting assumptions made on preference information underlying many standard datasets and experimental designs.

Moreover, with the promise of wearable and portable EEG devices, we foresee implications for engineering interactive systems that enable user models to be learned by directly observing the human cognitive system. These brain responses are unconstrained by the limitations of current user interfaces, interaction techniques, or requirements for users to explicitly express their reactions toward digital information. 

\section*{Acknowledgments}
We thank Zania Sovijärvi--Spapé for assistance in running the experiments and the members of the Cognitive Computing Group for discussions and support.

\bibliographystyle{plain}
%\bibliography{sample-bibliography}
\bibliography{bibliography}

%[{\includegraphics[width=1in,height=1.25in,clip,keepaspectratio]{mshell}}]
\begin{IEEEbiographynophoto}{Keith M. Davis III} is a PhD student at the University of Helsinki. He specializes in machine learning and collaborative brain-computer interfaces. 
\end{IEEEbiographynophoto}

\begin{IEEEbiographynophoto}{Michiel Spapé} received his PhD in Psychology from Leiden University in 2009. After working as a postdoc (Nottingham, Helsinki), and lecturer (Liverpool), he is now a docent in cognitive neuroscience at Helsinki University, focusing on emotion, perception/action, and EEG.
\end{IEEEbiographynophoto}

\begin{IEEEbiographynophoto}{Tuukka Ruotsalo} is an Associate Professor at the University of Copenhagen and an Academy Research Fellow at the University of Helsinki. He is an expert on cognitive computing, human-computer interaction, and brain-computer interfaces. \end{IEEEbiographynophoto}
\vfill
s
\end{document}